# Assessing the Impact of Electric Vehicle Charging on Residential Distribution Grids

Elias Raffoul
Department of Electrical and Computer Engineering
University of Houston
Houston, TX, USA
ejraffoul@uh.edu

Xingpeng Li
Department of Electrical and Computer Engineering
University of Houston
Houston, TX, USA
xli82@uh.edu

*Abstract*—To achieve net-zero carbon emissions, electrification in the transportation sector plays an important role. Significant increase of electric vehicles (EV) has been observed nationally and globally. While the transition to EVs presents substantial environmental benefits, it would lead to several challenges to the power grid due to EV charging activities. Growing EVs greatly increase peak loads on residential grids, particularly during evening charging periods. This surge can result in operational challenges, including greater voltage drops, increased power losses, and potential overloading violations, compromising grid reliability and efficiency. This study focuses on determining ampacity violations, and analyzing line loading levels in a 240-bus distribution system with 1120 customers, located in the Midwest U.S. By simulating a range of charging scenarios and evaluating EV chargers with varying power capacities under different distribution system voltage levels, this research aims to identify lines at risk of ampacity violations for various EV charging penetration rates up to 100%. The findings will provide valuable insights for utilities and grid operators, informing strategies for voltage level adjustments and necessary infrastructure reinforcements to effectively accommodate the growing energy demands associated with widespread EV adoption.

## I. Introduction

Electric vehicles (EVs) are emerging as a key solution to reduce greenhouse gas emissions from the transportation sector. According to the International Energy Agency's (IEA) "Net Zero Emissions by 2050 Scenario" report, the shift to electric mobility is essential for decarbonizing transportation, which remains a major driver of global emissions [1]. In the US, vehicles are major sources of $CO_2$ emissions, with transportation responsible for about 29% of the nation's greenhouse gas output. Within this sector, passenger cars and light-duty vehicles account for around 58% of emissions [2].

This growing demand for lower-emission alternatives is reflected in global EV sales, which are hitting record highs. BloombergNEF reports that 14% of new vehicle sales in 2023 were electric—double the share in 2021. Projections suggest EVs could comprise 35-40% of global car sales by 2030, potentially reaching 73% by 2040 [3]. Notably, if the largest automakers meet their targets, IEA predicts over 40 million electric cars could be sold annually by 2030, amplifying the need for infrastructure readiness. Policy support will be critical to accelerating EV adoption, particularly in addressing challenges related to battery supply chains, and charging networks as EV technology continues to mature [4].

While the rise of EVs offers a promising path to reducing carbon emissions, it also presents challenges to residential grids. The surge in EV adoption reshapes household energy demands and complicates grid management, especially during evening and nighttime charging hours. Electricity demand from road transport is projected to reach 8.3 petawatt hours by 2050 in the Net Zero Scenario [3]-[4], creating operational challenges such as voltage drops, increased power losses, and the risk of overloads, all of which can compromise grid reliability [5]. Studies suggest that integrating controlled charging strategies, particularly those that account for grid constraints, can alleviate strain by distributing peak loads more evenly, enhancing grid reliability [6].

While research provides valuable insights into voltage-related impacts, most studies focus on voltage drops and power quality issues, with fewer addressing line violations or ampacity limits—key indicators of grid stress that affect infrastructure longevity and safety. Some probabilistic studies assess the impact of high EV penetration on transformer loading and voltage levels, considering variables like charger ratings and user behaviors, but often overlook line loading violations [7]-[8]. Additionally, many studies focus on single or low-voltage systems, neglecting the response of medium-voltage residential networks, common in suburban areas, under varying voltage conditions [9]. Moreover, studies modeling EV penetration levels often use limited ranges, focusing on fixed or narrow adoption rates rather than exploring scenarios from 0% to 100% penetration [10]. Similarly, few studies conduct sensitivity analyses on varying charging capacities to understand their impact on grid performance under different loading conditions [11]. These gaps highlight the need for more comprehensive approaches that examine a wider range of EV adoption scenarios and charger capacities.

In contrast to prior studies, this paper analyzes a real-world 240-bus residential distribution system in the Midwest U.S. serving 1,120 customers. It uniquely models the impact of EV charging across multiple typical distribution voltage levels—4.16 kV, 6.9 kV, 13.8 kV, 23.9 kV, and 34.5 kV—commonly used in medium-voltage residential networks as specified in IEEE Std 141-1993 [12]. The work also investigates the impact of different EV penetration rates from 0% to 100% in 20% increments and examines various Level-2 EV charging capacities (5, 10, & 15 kW). By incorporating these elements, this study addresses critical gaps in prior research, offering an in-depth assessment of line loading levels, ampacity violations, and charger power sensitivity. These additions provide insights into potential overload risks unique to medium-voltage networks and practical guidance for utilities in establishing safe thresholds for EV integration. This comprehensive approach offers actionable recommendations for grid operators and policymakers to support widespread EV adoption while maintaining grid stability and infrastructure resilience.

## II. METHODOLOGY

To effectively analyze the impact of EV charging on the residential distribution grid, a systematic approach was developed. The methodology involves several key steps, beginning with data collection and analysis of residential load patterns, followed by the modeling of EV adoption scenarios at varying penetration levels. Each step in the process is designed to capture critical factors, such as load distribution across buses, voltage level settings, and the integration of EV charger capacities, which influence the overall power demands and operational limits of the grid. The flowchart in Fig. 1 illustrates the sequential approach used in this study, providing a clear visual representation of the data processing, simulation, and analysis stages to evaluate grid performance under various EV charging conditions.

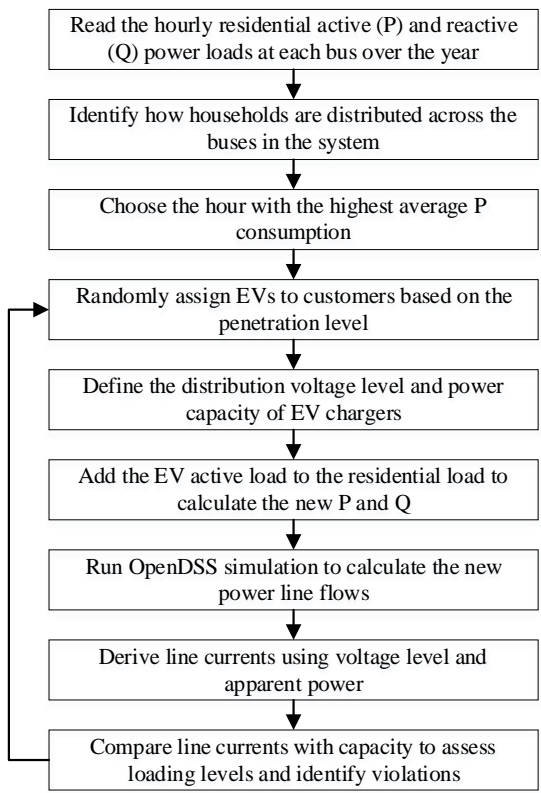

Fig. 1: Flowchart of the Proposed Methodology Procedure.

### A. System & Dataset Description

For our case study, we use a 69-kV substation system with one year of smart meter data from 2017 [13]. The system consists of three feeders, 240 primary network buses, spans 23 miles of primary feeder conductor, and serves 1,120 customers in the Midwest U.S. All customers are equipped with smart meters measuring hourly energy consumption (kWh), and the data comprises aggregated hourly consumption at each primary node, with time-aligned readings from customers connected to the same node summed to yield nodal aggregate consumption. The system includes standard electrical components such as overhead lines, underground cables, substation transformers with LTC, line switches, capacitor banks, and secondary transformers. The 240-node system is divided into Feeders A, B, and C, serving 17, 60, and 162 nodes, respectively.

Using the hourly energy consumption, active (kW) and reactive power (kVAR) are derived for spot loads. The hourly average kW demand (P) is estimated by assuming constant customer demand within each one-hour interval [14]. Reactive power (Q) is calculated by assigning a randomly selected power factor (pf) from the range of 0.9 to 0.95 for each customer. The resulting dataset includes hourly measurements of P and Q power for each node across all feeders over the year.

To estimate household distribution, the total annual active power across all feeders is calculated at 13,123,541 kWh. Dividing this by the 1,120 households served yields an average annual active power consumption per household of 11,717 kWh, translating to an average hourly consumption of approximately 1.34 kWh. This average was used to estimate the number of households at each bus by dividing each node's total power by the per-household average, rounded to the nearest whole number. The initial calculation yielded 1,125 households, which was adjusted to 1,120 by reducing one household from nodes with the highest counts.

To simulate worst-case loading conditions, the hour of peak average P consumption, occurring at 13:00 on July 12, 2017, was identified. Fig. 2 illustrates P and Q power profiles over 8 households at Bus 1004 on Feeder A. This load profile reflects typical residential behavior, with peak consumption occurring between 9:00 AM and 6:00 PM, particularly during summer months when cooling demand is higher.

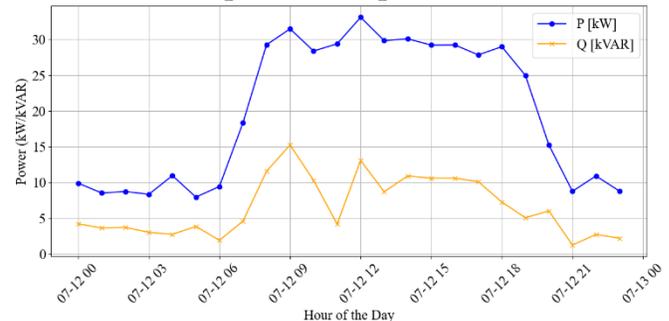

Fig. 2: 24-Hour Active and Reactive Powers for Bus 1004 on 2017-07-12.

### B. EV Charger Allocation Using the Mersenne Algorithm

To model the impact of EV adoption at varying penetration levels, EVs were allocated across buses based on five adoption rates: 20%, 40%, 60%, 80%, and 100%. The total number of EVs for each rate was determined as a percentage of the total number of homes in the network, and the EVs were then distributed probabilistically across the buses. The likelihood of each bus receiving EVs was proportional to the number of homes it served, ensuring that buses with more households had a higher probability of receiving more EVs. This approach mirrored the realistic distribution of EVs based on household density, while enforcing constraints to prevent assigning more EVs to a bus than the number of homes it serves. The allocation was performed using a random function from NumPy, which allows for weighted random selection. This function enables the probabilistic distribution of EVs across buses, with the probability of each bus being selected proportional to its share of the total households. The function relies on the Mersenne Twister algorithm, a widely used pseudorandom number generator that provides high-quality randomness. It starts with an initial "seed" value and applies this relation repeatedly to produce a sequence that appears random. One of its key strengths is its long period, ensuring that the numbers generated are diverse and not prone to short-term patterns [15]. This long period, combined with its high speed and statistical properties, makes the Mersenne Twister ideal for generating large volumes of random numbers efficiently.

*C. EV Load Estimation Procedure*

There are different levels of EV chargers, each designed to meet specific charging needs. These levels are classified as Level 1, Level 2, and DC fast chargers. Level 1 chargers typically use a standard 120V outlet, providing 1.2 kW of power, making them suitable for overnight charging at home. Level 2 chargers operate at 240V and offer charging power between 3.3 kW and 19.2 kW, providing faster charging, typically installed in residential or commercial settings. DC fast chargers, operating at higher voltages, can provide charging power from 25 kW to 350 kW, enabling rapid charging at public stations [16]. In our study, we focus on Level 2 EV chargers with capacities of 5 kW, 10 kW, and 15 kW, as these are commonly used in residential settings. These chargers are distributed across the system, and their impact on the grid is modeled by updating P and Q power values at each bus. P is increased by the charging power capacity of each EV, assuming the simulation is run for one hour. For instance, with a 10-kW charger, the active power increases by 10 kW per EV, and the updated value is computed accordingly. Once P is updated, the next step is to recalculate Q. With the updated active and reactive power values, the loads in OpenDSS, an open-source software used for power flow simulations, are adjusted. The power flow simulation is then run to obtain the line power values, allowing us to assess the grid's performance under varying levels of EV adoption.

*D. Line Current & Loading Level Calculation*

The system comprises several types of lines, each defined by its configuration (overhead (OH) or underground (UG)) and the number of phases (1-phase, 2-phase, or 3-phase). These lines have varying current capacities per phase, ranging from 242 A to 357 A. The apparent power (S), derived from P and Q, is used with the system's voltage and phase configuration to compute line currents. For three-phase systems, the current is calculated based on the line-to-line voltage, while for single-phase systems, the calculation uses the line-to-neutral voltage. To assess the loading levels and potential violations in the system, the line loading is computed as the ratio of the actual current to the line's rated current as formulated in (1).

$$Line\ Loading = \frac{I_{actual}}{I_{rated}} \qquad (1)$$

$$Violation\ \% = \frac{I_{actual} - I_{rated}}{I_{rated}} * 100 \qquad (2)$$

If the line loading exceeds 100%, a violation occurs, indicating that the line is carrying more current than its rated capacity. The severity of the violation is quantified by calculating the violation percentage, which is determined by (2). This percentage reflects the extent to which the line is overloaded. A higher violation percentage indicates a greater deviation from the rated capacity. This calculation helps identify areas where the grid may be overburdened and requires corrective action to prevent potential failures or damage.

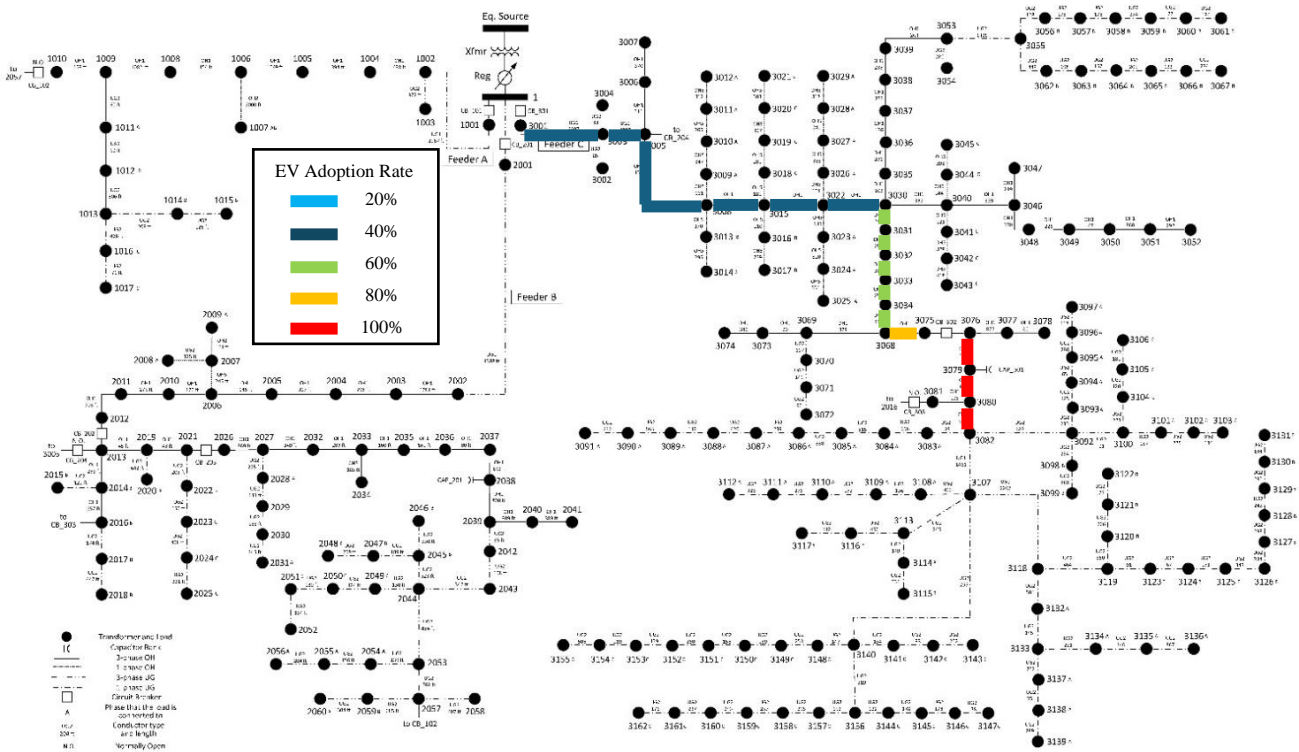

Fig. 3: Line Violations for different EV penetration rates given distribution voltage level of 6.9kV, assuming the charging power is 10kW.

## III. RESULTS

After calculating line currents across various voltage levels and EV penetration rates, we highlighted the line violations in Fig. 3 for a system voltage of 6.9 kV and an EV charger capacity of 10 kW. This figure provides a visual representation of the system and the identified violations, adapted from a figure in [13]. A cumulative color-coding approach is employed in the legend to represent these violations at different EV adoption rates. For instance, light blue marks lines that experience violations at a 20% EV rate, while teal indicates additional lines that become overloaded at 40% penetration, also encompassing those previously overloaded at 20%. This pattern continues, with each subsequent color representing an increasing penetration level and the cumulative effect on the network. The illustration reveals that lines closest to the substation are most vulnerable to overloading, as they carry the highest cumulative power flow from downstream buses. As EV penetration increases, the stress extends progressively further along the network, with each subsequent line having a higher chance of being overloaded. This cascading effect emphasizes the strain on distribution systems as EV adoption grows.

TABLE I.  LOADING LEVEL STATISTICS FOR EV CHARGER P = 10 KW

| Voltage Level | Loading Level Metric | EV Adoption Rate | | | | | |
|---|---|---|---|---|---|---|---|
| | | 0% | 20% | 40% | 60% | 80% | 100% |
| 4.16 kV | Min % | 0.03 | 0.03 | 0.03 | 0.03 | 0.03 | 0.03 |
| | Max % | 97.5 | 162.5 | 226.7 | 303.1 | 357.6 | 402.9 |
| | Avg % | 11.3 | 19.2 | 26.6 | 34.7 | 40.8 | 46.5 |
| 6.9 kV | Min % | 0.02 | 0.02 | 0.02 | 0.02 | 0.02 | 0.01 |
| | Max % | 58.7 | 101.4 | 138.3 | 184.0 | 211.7 | 243.1 |
| | Avg % | 6.8 | 11.6 | 16.5 | 20.8 | 24.9 | 28.1 |
| 13.8 kV | Min % | 0.01 | 0.01 | 0.01 | 0.01 | 0.01 | 0.01 |
| | Max % | 29.3 | 48.5 | 67.2 | 91.4 | 105.6 | 121.7 |
| | Avg % | 3.4 | 5.8 | 8.3 | 10.5 | 12.4 | 14.0 |
| 23.9 kV | Min % | 0.01 | 0.00 | 0.00 | 0.00 | 0.00 | 0.00 |
| | Max % | 17.0 | 28.7 | 38.4 | 53.3 | 60.2 | 70.2 |
| | Avg % | 1.9 | 3.3 | 4.7 | 6.1 | 7.1 | 8.1 |
| 34.5 kV | Min % | 0.00 | 0.00 | 0.00 | 0.00 | 0.00 | 0.00 |
| | Max % | 11.8 | 20.4 | 28.3 | 36.4 | 42.7 | 48.6 |
| | Avg % | 1.4 | 2.4 | 3.2 | 4.3 | 5 | 5.6 |

Table I shows loading level statistics at various voltage levels for different EV adoption rates with a 10-kW charger. At 4.16 kV, the maximum loading percentage increases sharply from 97.5% with no EVs to over 400% at full adoption, signaling a high overload risk. The average loading also rises, reflecting increased system strain. At higher voltages, such as 13.8 kV and 34.5 kV, the system's resilience improves, with lower loading percentages—maximum loading at 13.8 kV is 121.7% and at 34.5 kV is 48%. This highlights the growing strain at higher EV adoption rates and the improved capacity at higher voltage levels.

Table II examines line violations across voltage levels and EV adoption rates with 10 kW chargers, where any loading exceeding 100% in Table I is considered a violation. At 4.16 kV, violations worsen with higher EV adoption, reaching 301.7% at full adoption, with violations rising from 6 to 38 instances. At 6.9 kV, violations also increase, from 37.8% at 40% adoption to 142.7% at full adoption. The 13.8 kV level shows fewer violations, peaking at 21.5% at full adoption, while 23.9 kV and 34.5 kV show no violations, indicating better capacity for handling EV loads. This reinforces the trend that higher voltage levels offer stronger infrastructure for EV integration. Table III reveals the critical EV adoption thresholds for triggering line violations, with a 10-kW charger: 2% for 4.16 kV, 21% for 6.9 kV, and 71% for 13.8 kV.

TABLE II.  LINE VIOLATIONS STATISTICS FOR EV CHARGER P = 10 KW

| Voltage Level | Violation Metric | EV Adoption Rate | | | | |
|---|---|---|---|---|---|---|
| | | 20% | 40% | 60% | 80% | 100% |
| 4.16 kV | Count | 6 | 14 | 27 | 37 | 38 |
| | Min % | 20.9 | 1.3 | 2.9 | 5.2 | 11.4 |
| | Max % | 59.7 | 127.6 | 198.9 | 249.1 | 301.7 |
| | Avg % | 42.4 | 57.4 | 66.1 | 72.8 | 92.1 |
| 6.9 kV | Count | 0 | 6 | 11 | 13 | 19 |
| | Min % | | 4.0 | 0.3 | 9.9 | 0.8 |
| | Max % | | 37.8 | 80.9 | 108.6 | 142.7 |
| | Avg % | | 22.0 | 33.9 | 48.0 | 49.1 |
| 13.8 kV | Count | 0 | 0 | 0 | 2 | 4 |
| | Min % | | | | 5.4 | 4.2 |
| | Max % | | | | 7.7 | 21.5 |
| | Avg % | | | | 6.5 | 13.3 |
| 23.9 kV | Count | 0 | | | | |
| 34.5 kV | Count | 0 | | | | |

TABLE III.  EV RATES TRIGGERING LINE VIOLATIONS

| Voltage Level | 4.16 kV | 6.9 kV | 13.8 kV |
|---|---|---|---|
| EV Rate | 2% | 21% | 71% |

A sensitivity analysis of EV charger power levels (5 kW, 10 kW, and 15 kW) shows that higher-power chargers lead to more line violations as adoption increases. Fig. 4 shows 15 kW chargers cause 38 violations at 60% adoption, while 5 kW chargers result in 23 violations at 100% adoption. Fig. 5 reveals that higher charger power and adoption increase loading levels, with maximum loading reaching 396.3% at 15 kW and 100% adoption, compared to 301.7% at 10 kW.

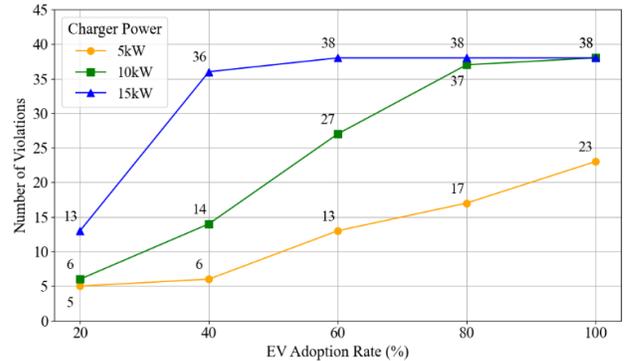

Fig. 4: Line Violation Count for Varying Power Capacities at 4.16 kV.

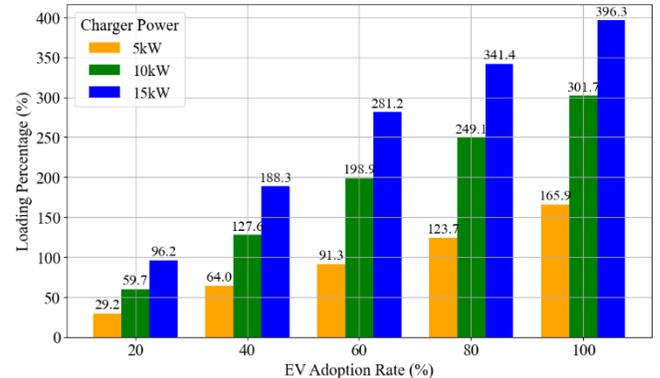

Fig. 5: Max Loading Percentage for Varying Power Capacities at 4.16 kV.

As shown in Fig. 2, power consumption is lower from 6:00 PM to 9:00 AM, with the lowest average consumption across buses occurring at 4:00 AM. After running a simulation with 4:00 AM base load conditions, Figs. 6 and 7, reveal that increasing EV adoption raises voltage levels more at lower voltage lines. At 100% EV adoption, 4.16 kV lines reach 46.5% loading during peak hours, compared to 40.2% off-peak. At higher voltage levels (23.9 kV and 34.5 kV), the differences are smaller, and average loading stays below 10%.

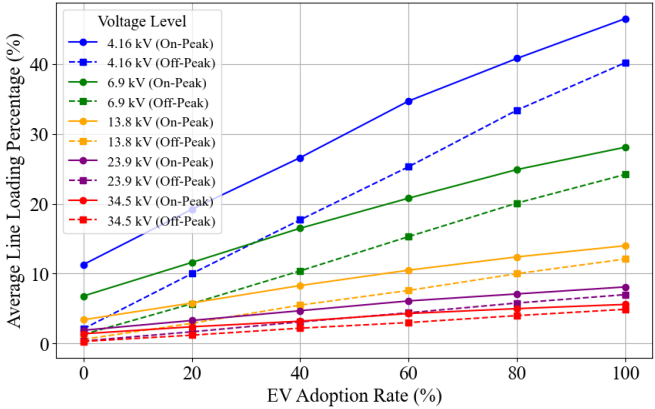

Fig. 6: Average Line Loading vs. EV Adoption Rate for On-Peak and Off-Peak Hours under Different Voltage Levels.

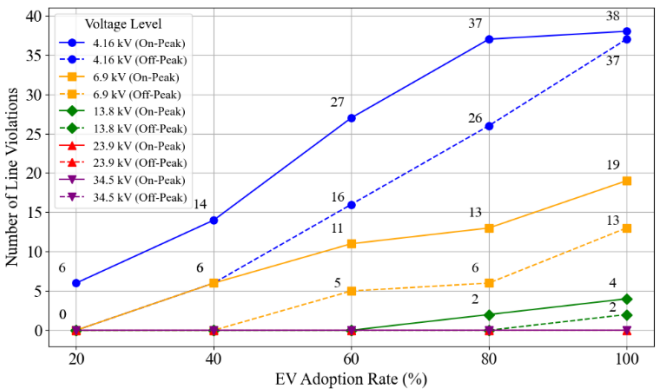

Fig. 7: Number of Line Violations vs. EV Adoption Rate for On-Peak and Off-Peak Hours under Different Voltage Levels.

Fig. 7 reinforces the findings by highlighting line violations, with 4.16 kV experiencing the highest number of violations—14, 27, and 37—at 40%, 60%, and 80% adoption rates respectively during peak hours, compared to 6, 16, and 26 during off-peak hours. Both figures consistently show reduced impacts during off-peak times, underscoring the importance of time-of-use considerations in EV integration planning.

## IV. CONCLUSION

The transition to EVs introduces challenges to distribution grid reliability and efficiency. This study examined the impact of varying EV adoption rates on cables with different voltage levels, focusing on sensitivity analysis on line loading and ampacity violations. Our work shows that higher EV penetration can stress distribution lines, leading to ampacity violations. At 4.16 kV, violations occur at just 2% EV adoption, while 13.8 kV networks can support up to 71%. Additionally, charging power plays a significant role—15 kW chargers cause more violations than 5 kW chargers, particularly at lower voltage levels.

Future work should explore solutions to ensure grid stability with increasing EV adoption, including voltage adjustments, capacity reinforcement, and controlled charging strategies. Specifically, upgrading lines near substations, promoting off-peak charging, and considering higher voltage levels in high-adoption areas could be cost-effective. Additionally, evaluating the economic trade-offs between these strategies, along with energy storage and smart charging, will be key to maintaining grid reliability.

5